\documentclass[reprint, superscriptaddress, aps, pra, tightenlines, floatfix]{revtex4-1}

\pdfoutput=1

\usepackage[utf8]{inputenc}
\usepackage[T1]{fontenc}
\usepackage[english]{babel}

\usepackage{amsmath}
\usepackage{amssymb}
\usepackage{amsfonts}
\usepackage{mathrsfs}
\usepackage{dsfont}
\usepackage{bm}% bold math

\usepackage[version=4]{mhchem} %for chemical formula
\usepackage{textgreek} %for greek
\usepackage{dcolumn}% Align table columns on decimal point

\usepackage{graphicx}
\usepackage[caption=false]{subfig}
\usepackage{enumerate}
\usepackage[shortlabels]{enumitem}

\usepackage{hyphenat}
\usepackage{hyperref}

\usepackage[capitalise]{cleveref}
\crefname{section}{Sec.}{Secs.}% APS style uses abbreviations
\Crefname{section}{Section}{Sections}
\usepackage{hypcap}
\usepackage{braket}
\usepackage{floatrow}
\usepackage{siunitx}
\newcommand{\psnspd}{P\nobreakdash-SNSPD}
\newcommand{\ketbra}[2]{{|{#1}\rangle\!\langle{#2}|}}
\newcommand{\tr}{\text{tr}}

\newcommand{\id}{\mathds{1}}

\begin{document}

% \preprint{APS/123-QED}

\title{Enhanced heralded single-photon source with a photon-number-resolving parallel superconducting nanowire single-photon detector}

\author{Lorenzo~Stasi}
\email[Corresponding author: ]{lorenzo.stasi@idquantique.com}
\affiliation{Department of Applied Physics, University of Geneva, CH-1211 Geneva, Switzerland}
\affiliation{ID Quantique SA, CH-1227 Carouge, Switzerland}

\author{Patrik~Caspar}
\affiliation{Department of Applied Physics, University of Geneva, CH-1211 Geneva, Switzerland}

\author{Tiff~Brydges}
\affiliation{Department of Applied Physics, University of Geneva, CH-1211 Geneva, Switzerland}

\author{Hugo~Zbinden}
\affiliation{Department of Applied Physics, University of Geneva, CH-1211 Geneva, Switzerland}

\author{F\'elix~Bussi\`eres}
\affiliation{ID Quantique SA, CH-1227 Carouge, Switzerland}
% \affiliation{Department of Applied Physics, University of Geneva, CH-1211 Geneva, Switzerland}

\author{Rob~Thew}
\affiliation{Department of Applied Physics, University of Geneva, CH-1211 Geneva, Switzerland}

\date{\today}% It is always \today, today,

\begin{abstract}
Heralded single-photon sources (HSPS) intrinsically suffer from multiphoton emission, leading to a trade-off between the source's quality and the heralding rate. A solution to this problem is to use photon-number-resolving (PNR) detectors to filter out the heralding events where more than one photon pair is created. Here, we demonstrate the use of a high-efficiency PNR superconducting nanowire single-photon detector (SNSPD) as a heralding detector for a HSPS. By filtering out higher-order heralding detections, we can reduce the $g^{(2)}(0)$ of the heralded single photon by $(26.6 \pm 0.2)\,\%$, or alternatively, for a fixed pump power, increasing the heralding rate by a factor of $1.363 \pm 0.004$ for a fixed $g^{(2)}(0)$. Additionally, we use the detector to directly measure the photon-number distribution of a thermal mode and calculate the unheralded $g^{(2)}(0)$. We show the possibility to perform $g^{(2)}(0)$ measurements with only one PNR detector, with the results in agreement with those obtained by more common-place techniques which use multiple threshold detectors. Our work shows that efficient PNR SNSPDs can significantly improve the performance of HSPSs and can precisely characterize them, making these detectors a useful tool for a wide range of optical quantum information protocols.
\end{abstract}

\maketitle

%%%%%%%%%%%%%%%%%%%%%%%%%%%%%%%%%%%%%%%%%%%%%%%%%%%%%%%%%%%%%%%%%%%%%%%%%%%%%%%%%%%%%%%%%%%%%%%%%%%%%%%%%%%%%%%%%%%%%%%%%%%%%%%%%%%%%%%%%%%%%%
\section{Introduction}
\label{sec:introduction}
Over the past decades, there have been remarkable developments in the field of quantum technologies. In particular, photonic systems employing single photons have been used in a variety of applications, ranging from quantum communication and repeater protocols~\cite{gisin2007quantum, sangouard2011quantum} to linear optical quantum computing and Gaussian boson sampling~\cite{knill2001scheme, kok2007linear, slussarenko2019photonic, spring2013boson, Bentivegna2015, zhong2018}.
A convenient and versatile tool to generate single photons are heralded single-photon sources (HSPS)~\cite{Castelletto2008, Eisaman2011}. They have the advantage of operating at room temperature, are wavelength and bandwidth tunable, and can produce indistinguishable and pure photons~\cite{mosley2008heralded, Bruno2014a, Graffitti2018independent}.

The photon generation mechanism in HSPS is, however, probabilistic and multi-photon events can occur. Such events are undesired, leading to a decrease in the single-photon fidelity \cite{Scott2020}. Therefore, to minimize their impact, HSPSs are often used in the low-squeezing regime ($\mu\ll1$, where $\mu$ is the mean photon-pair number per pulse). However, in such a regime the probability of emitting vacuum states increases, reducing the generation rate of single photons.

A possible solution to this problem is to use a photon-number resolving (PNR) detector as the heralding detector. In this way, it is possible to filter out events when more than one photon is detected, therefore lowering the multi-photon contribution in the heralded state. 
In principle, such a scheme would allow one to work with a higher $\mu$, making it possible to increase the single-photon generation rate, allowing for higher heralding rates.

Transition-edge sensors (TESs) have shown to be able to distinguish high numbers of photons with high probability~\cite{lita2008counting,morais2020precisely}. 
However, the long recovery time of several microseconds precludes their application in high rate experiments, limiting their operation to the kHz regime. In addition, TESs need ultra-low temperatures ($<100$~mK) which requires a complex cryogenic system.
%increases the complexity of the cryogenic system.

Recently, SNSPDs have demonstrated few-photons PNR capability based on the signal's slew rate~\cite{cahall2017multi} or on the voltage pulse amplitude when used in combination with an impedance matching taper~\cite{zhu2020resolving}. Both methods have been employed in recent experiments to improve the heralded $g^{(2)}(0)$ measurement of HSPSs~\cite{Sempere-Llagostera2022, Davis2021}. In the first approach, one has to fit the waveform's rising edge to retrieve the photon-number event, which requires detectors with very low timing jitter.
In the second one, the impedance matching taper acts as a kinetic inductive element, which increases the recovery time of the overall detector, potentially limiting its usage at high repetition rates. 

A promising alternative technology is based on SNSPDs in a parallel configuration (\psnspd{}) \cite{marsili2009superconducting, perrenoud2021operation, stasi2022high}. It consists of an array of several pixels (the single SNSPD element) which are electrically connected in parallel. With respect to an array of independent pixels, a \psnspd{} requires only one coaxial cable for the read-out, which is a practical advantage. In fact, information on the number of pixels that clicked can be extracted from the signal amplitude, which eases the photon-number discrimination. %not having to rely on the timing jitter.
In addition, each pixel is much shorter than a conventional SNSPD, thus it can recover faster thanks to the lower kinetic inductance. By combining this feature with a recently developed architecture~\cite{perrenoud2021operation}, which completely prevents electrical crosstalk and latching, the \psnspd{} can be used to measure photon-number statistics~\cite{stasi2022high} in high repetition rate regimes.

This paper reports on the improvements to a HSPS that can be achieved through use of a high-efficiency ($>\SI{80}{\%}$) \psnspd{}, when used to distinguish multi-photon detection events. We first introduce the theoretical tools that we use to describe the employed source and the detectors. 
Then we conduct two experiments, first we show the improvement from using the \psnspd{} as a heralding detector as part of a HSPS in terms of improving the $g^{(2)}(0)$ of the heralded single-photon state.
Second, we show that a \psnspd{} can be used to measure $g^{(2)}(0)$ on a single spatial mode. To this end, we characterize the unheralded output of a HSPS by reconstructing its thermal statistics. 

%%%%%%%%%%%%%%%%%%%%%%%%%%%%%%%%%%%%%%%%%%%%%%%%%%%%%%%%%%%%%%%%%%%%%%%%%%%%%%%%%%%%%%%%%%%%%%%%%%%%%%%%%%%%%%%%%%%%%%%%%%%%%%%%%%%%%%%%%%%%%%

\section{Theory}
\label{sec:theory}
The ideal state generated by a SPDC source is a two-mode squeezed vacuum (TMSV) state described by~\cite{Walls2008}
\begin{equation}
\label{eq:tmsv}
\begin{split}
    \ket{\Psi}_{si} &= \sqrt{1-\lambda^2} \sum_{n=0}^{\infty} \lambda^n \ket{nn}_{si} \\
    &= \sum_{n=0}^{\infty} \sqrt{\frac{\mu^n}{(\mu+1)^{n+1}}} \ket{nn}_{si},
\end{split}
\end{equation}
where $\lambda=\tanh{r}$ with the squeezing parameter $r$ and the mean photon number $\mu=\sinh{r}^2$. Note that the marginal states of the signal ($s$) and idler ($i$) modes are thermal states with photon-number probability distribution 
\begin{equation}
\label{eq:thermaldist}
    p_n = \frac{\mu^n}{(\mu+1)^{n+1}}.
\end{equation}
First, we consider the scenario shown in \cref{fig:setup}a, in which the signal photon is sent to a \psnspd{} acting as a heralding detector, $\mathrm{D_h}$, and the idler photon is sent to a 50/50 beam splitter followed by two threshold SNSPDs, $\mathrm{D_a}$ and $\mathrm{D_b}$. This allows evaluation of the second-order auto correlation function~\cite{Walls2008} 
\begin{equation}
\label{eq:g2}
    g^{(2)}(0) = \frac{\langle \hat{n}(\hat{n}-1) \rangle}{\langle \hat{n} \rangle ^2} = \frac{\sum_n n(n-1)p_n}{\big(\sum_n n p_n\big)^2},
\end{equation}
where $\hat{n}$ is the photon-number operator and $p_n$ the photon-number probability distribution of the state. An ideal single-photon state exhibits $g^{(2)}(0)=0$ and a thermal state fulfills $g^{(2)}(0)=2$.

In order to obtain analytical equations describing the single and coincidence detection probabilities per pump pulse, we use the approach of Ref.~\cite{Takeoka2015}. In this formalism, the TMSV state $\rho=\ketbra{\Psi}{\Psi}_{si}$ can be expressed by a $4\times 4$ covariance matrix with $\mu$ as a single free parameter. Furthermore, the action of beam splitters can be modeled by Gaussian unitary operations. The formalism additionally allows modes to be traced out, and so we are able to model transmission loss on a given mode by introducing an auxiliary mode, applying a beam splitter operation between the two modes, and finally tracing out the auxiliary mode. Moreover, the formalism also allows for calculation of the expectation value of a given Gaussian state after projection onto vacuum. Therefore, we can model threshold detectors described by positive-operator-valued measure (POVM) elements $E_0=\ketbra{0}{0}$ corresponding to a no-click outcome and $E_c=\id - \ketbra{0}{0}$ to a click outcome. To obtain a model for the \psnspd{}, we note that each of the $N$ pixels of the detector is a threshold detector. Therefore, a physically intuitive model for the \psnspd{} consists of a sequence of beam splitters, with splitting ratios corresponding to the characterized pixel efficiencies, and $N$ threshold detectors (see Appendix A for further details).

To additionally take account of the non-unit spectral purity of our source, we assume a multimode state $\rho=\varrho_1\otimes\dots\otimes\varrho_M$. Here $\varrho_k= \ketbra{\Psi}{\Psi}_{k}$, as defined in \cref{eq:tmsv}, describes a single Schmidt mode with mean photon number $\lambda_k\mu$, where the Schmidt coefficients are normalized such that $\sum_k\lambda_k=1$. In the case of a SPDC source, the Schmidt coefficients can be estimated by a measurement of the joint spectral intensity~\cite{Zielnicki2018}. In this way, we obtain a model to accurately describe the expected single and coincidence detection probabilities per pump pulse and the heralded $g^{(2)}(0)$
\begin{equation}
\label{eq:g2cond}
    g_{\mathrm{h}}^{(2)}(0) \approx \frac{p_\mathrm{h}p_\mathrm{hab}}{p_\mathrm{ha}p_\mathrm{hb}},
\end{equation}
where $p_\mathrm{h}$ is the probability of a heralding detection, $p_\mathrm{ha(b)}$ the probability of a coincidence detection between detectors $\mathrm{D_h}$ and $\mathrm{D_a}$~($\mathrm{D_b}$), and $p_\mathrm{hab}$ the triple-coincidence probability between $\mathrm{D_h}$, $\mathrm{D_a}$, and $\mathrm{D_b}$. For details on the derivation and the explicit formulas for the probabilities, see Appendix~\ref{sec:app_model}.

Similarly, the unconditional $g^{(2)}(0)$ on one spatial mode of the TMSV state can be evaluated as
\begin{equation}
\label{eq:g2uncond}
    g_\mathrm{unc.}^{(2)}(0) \approx \frac{p_\mathrm{ab}}{p_\mathrm{a} p_\mathrm{b}},
\end{equation}
where $p_\mathrm{a(b)}$ is the probability of a detection on $\mathrm{D_a}$~($\mathrm{D_b}$) and $p_\mathrm{ab}$ the probability of a coincidence detection between $\mathrm{D_a}$ and $\mathrm{D_b}$. We note that in the case where we have direct access to the photon-number probability distribution $p_n$ of the state, e.g. by measuring it with a PNR detector, one can also use Eq.~\eqref{eq:g2} to obtain $g^{(2)}(0)$.

%%%%%%%%%%%%%%%%%%%%%%%%%%%%%%%%%%%%%%%%%%%%%%%%%%%%%%%%%%%%%%%%%%%%%%%%%%%%%%%%%%%%%%%%%%%%%%%%%%%%%%%%%%%%%%%%%%%%%%%%%%%%%%%%%%%%%%%%%%%%%%

\section{Experiment}
\label{sec:experiment}
\begin{figure}
    \capstart
    \centering
    \includegraphics[width = 1.0\columnwidth]{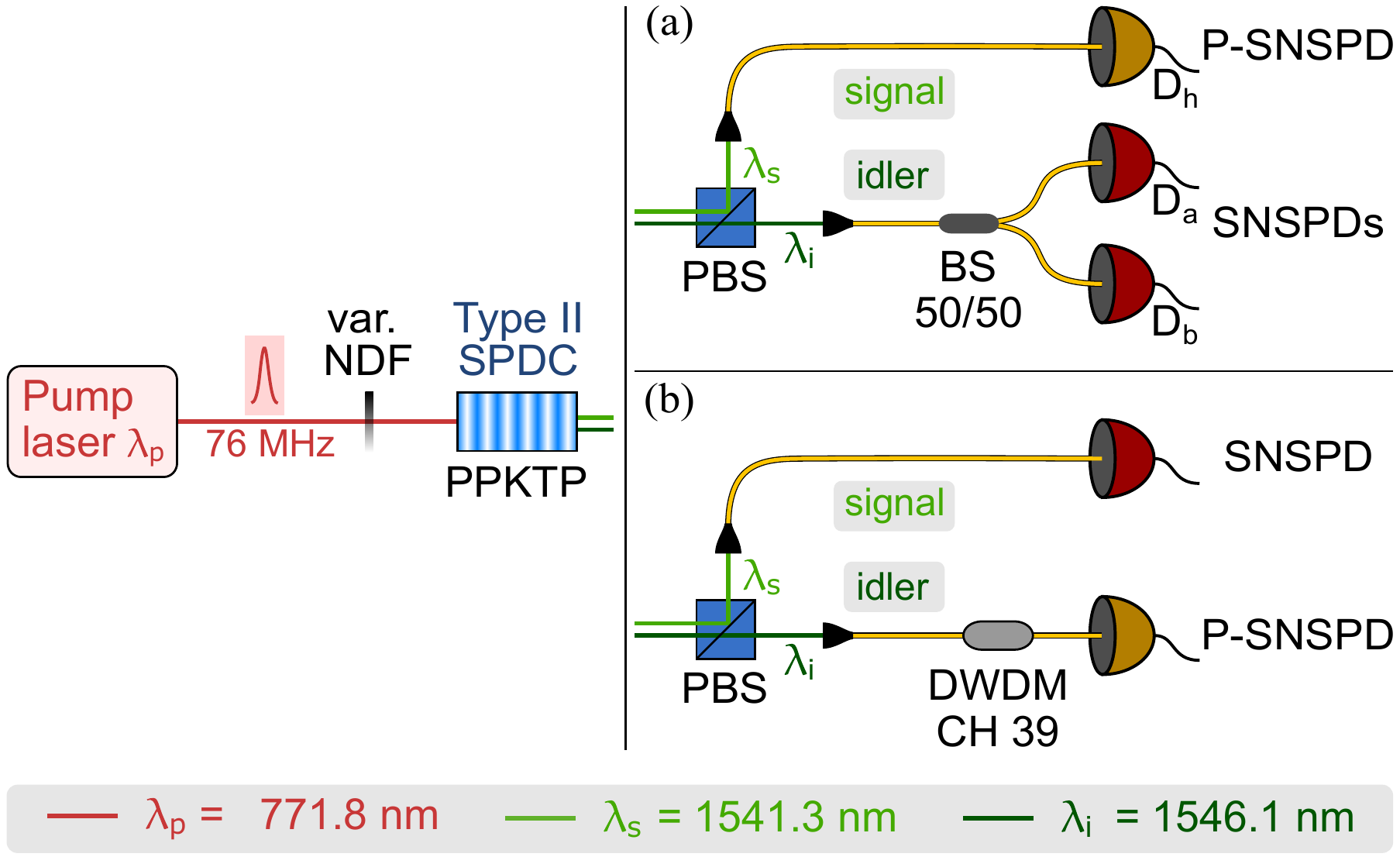}
    \caption{Experimental setup. A Ti:sapphire laser in the picosecond pulsed regime at $\lambda_p = \SI{771.8}{nm}$ with a repetition rate of \SI{76}{MHz} is used to pump a \SI{30}{mm} long periodically poled potassium titanyl phosphate (PPKTP) bulk nonlinear crystal with poling period $\Lambda=\SI{46.2}{\mu m}$. In this way, non-degenerate signal ($\lambda_s = \SI{1541.3}{nm}$) and idler ($\lambda_i = \SI{1546.1}{nm}$) photon pairs are generated via type-II SPDC, where the pump power can be varied with a reflective variable neutral density filter (NDF). Signal and idler photons are separated by a polarizing beam splitter (PBS) and coupled into single-mode optical fibers. (a) The heralding signal photons are detected by the {P-SNSPD} while the idler photons are sent to a 50/50 beam splitter (BS) and detected by threshold SNSPDs. (b) The idler photons are spectrally filtered by a dense wavelength division multiplexer (DWDM) channel 39 and detected by the {P-SNSPD} for the reconstruction of the thermal photon number statistics.}
    \label{fig:setup}
\end{figure}

In a first experiment, as illustrated in Fig.~\ref{fig:setup}a, we employ a \psnspd{} as the heralding detector in a HSPS and assess its performance by measuring $g_{\mathrm{h}}^{(2)}(0)$ for different pump power settings. We use the \psnspd{} in two different configurations: as a threshold detector, where all the $(n\geq1)$-click events are considered as a (single) detection, and as a PNR detector, where only $(n=1)$-click events are considered.
%without discarding $n\geq2$-click events, and as a PNR detector, where $n\geq2$-click events are discarded. 

In a second experiment, shown in \cref{fig:setup}b, we replace the 50/50 beam splitter and the two standard SNSPDs by a single \psnspd{} to reconstruct the photon-number probability distribution. Using \cref{eq:g2} we calculate $g_\mathrm{unc.}^{(2)}(0)$ and compare it to the values obtained when using the more standard method, where a 50/50 beam splitter and two threshold detectors are used, via \cref{eq:g2uncond}. Here, we additionally filter the idler mode with a dense wavelength division multiplexer (DWDM) in order to herald spectrally pure photons and suppress leaking signal photons due to the finite extinction ratio of the polarizing beam splitter which separates the signal and idler photons.

The spectral purity of the heralded single photons is characterized by a joint spectral intensity measurement~\cite{Zielnicki2018,Graffitti2018design} and was found to be $\sim\SI{84}{\%}$. The total transmission and detection efficiencies for signal and idler are measured at low pump power ($\mu\approx 5\times 10^{-4}$) using the method described in~\cite{Klyshko1980} and amount to $\eta_s=(C_\mathrm{ha}+C_\mathrm{hb})/(C_\mathrm{a}+C_\mathrm{b})=\num{0.6348(5)}$ and $\eta_i=(C_\mathrm{ha}+C_\mathrm{hb})/C_\mathrm{h}=\num{0.6051(6)}$, where $C$ denotes the (coincidence) counts on the corresponding detectors within \SI{3}{min} of integration time. 

The two standard threshold detectors are in-house developed single-pixel SNSPDs made from molybdenum silicide (MoSi) and have system detection efficiencies of about \SI{85}{\%} and \SI{83}{\%}, respectively.
The \psnspd{} is a four-pixel MoSi SNSPD where the amplitude of the electrical readout signal is dependent on the number of pixels that click in the detection process. To avoid electrical and thermal crosstalk between the pixels, we employed the architecture developed in Ref.~\cite{perrenoud2021operation}. 
In order to characterize the \psnspd{}, light with known statistics (Poissonian with $\mu=1$, in our case) is sent onto the detector. The photon-counting statistics are collected and, by employing an optimization algorithm, one can obtain the conditional probabilities matrix $\mathbf{P}$. Its elements ${P_{nm}}$ denote the probabilities of registering an $n$-click event when $m$ photons are incident on the detector. Therefore, an initial photon-number probability distribution $p_m$ leads to an $n$-click probability recorded by the \psnspd{} of $q_n = \sum_{m=0}^{\infty} P_{nm} \, p_m$. 
Even though $m$ can go to infinity, practically it is stopped at a finite value $M$. Therefore, $\mathbf{P}$ has dimension $(N+1)\times M$, where $N$ is the number of pixels of the \psnspd{}. By inverting $\mathbf{P}$, the incident photon-number probability distribution $p_m$ can be reconstructed from the detected click probability distribution $q_n$. In our case for $N=4$ and $M=9$, we measured $P_{11}=84\%$, $P_{12}=55\%$, $P_{22}=42\%$, $P_{13}=31\%$ and $P_{14}=17\%$. %data from 2022-03-30
For further information, see Ref. \cite{stasi2022high}.

\begin{figure}[t]
    \capstart
    \centering
    \includegraphics[width = 1.0\columnwidth]{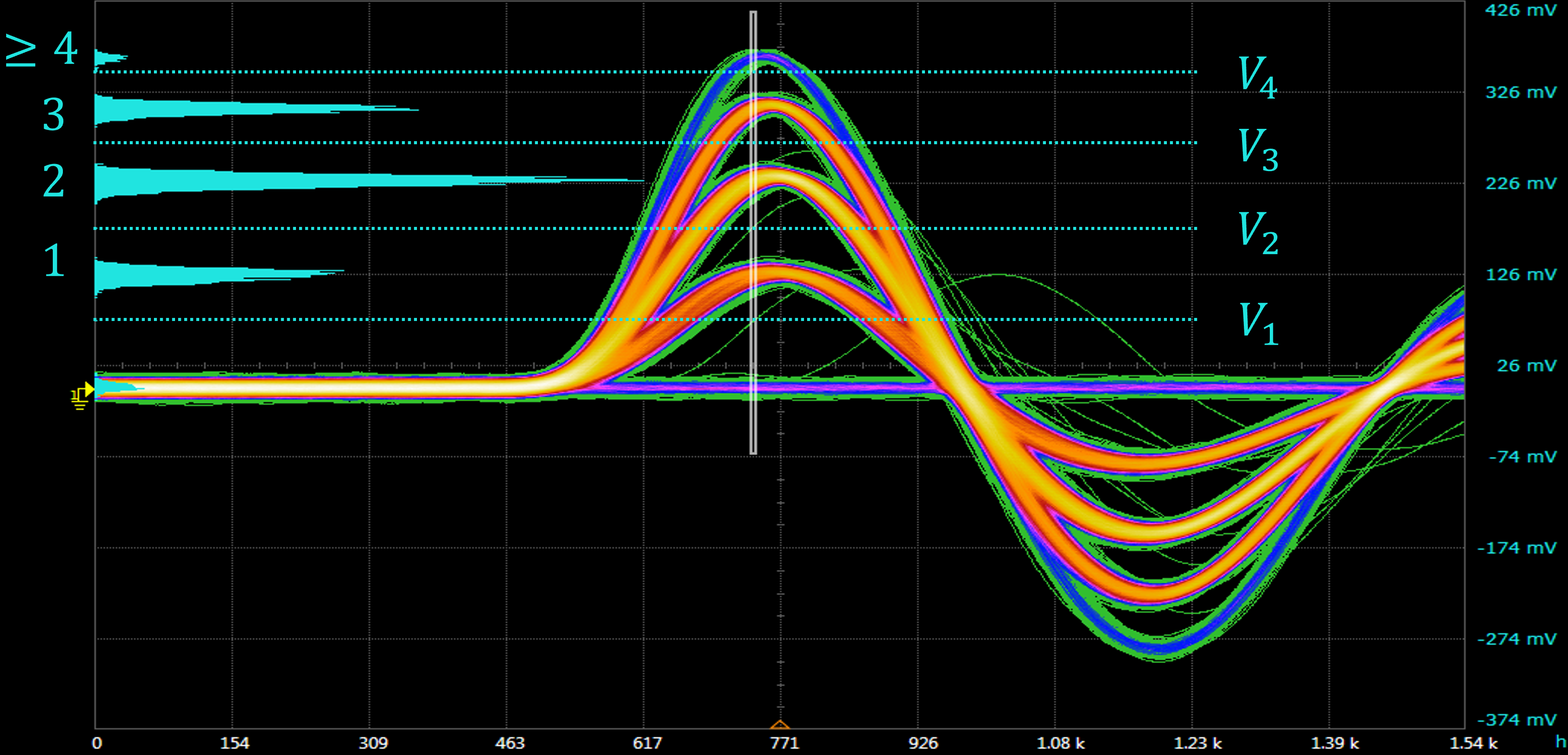}
    \caption{Oscilloscope persistence traces of the electrical signals generated by the {P-SNSPD}. On the left, are reported the waveforms histograms corresponding to each $n$-click event, taken from the vertical white slice. The $n$-click events are discriminated by setting different voltage thresholds on the time tagger. 
    }
    \label{fig:traces}
\end{figure}

For the photon-number discrimination, the electrical readout signal of the \psnspd{} is separated in two by a coaxial power splitter and discriminated by a time controller (ID Quantique ID900) at two different voltage thresholds, corresponding to a detection of $n\geq 1$ photons and $n\geq 2$ photons (see \cref{fig:traces}).
In the second experiment, where the photon-number probability distribution is reconstructed, we also use a third discrimination level corresponding to a detection of $n\geq 3$ photons. 
The time controller additionally takes the electrical pickup signal from the pump laser and the readout signals from the threshold SNSPDs. All detection events are taken within a \SI{1}{ns} time window with respect to the pump pulses to reduce dark count contributions, and their timestamps are saved for the data analysis.

%%%%%%%%%%%%%%%%%%%%%%%%%%%%%%%%%%%%%%%%%%%%%%%%%%%%%%%%%%%%%%%%%%%%%%%%%%%%%%%%%%%%%%%%%%%%%%%%%%%%%%%%%%%%%%%%%%%%%%%%%%%%%%%%%%%%%%%%%%%%%%

\section{Results}
\label{sec:results}

\begin{figure}
    \capstart
    \centering
    \includegraphics[width=\columnwidth]{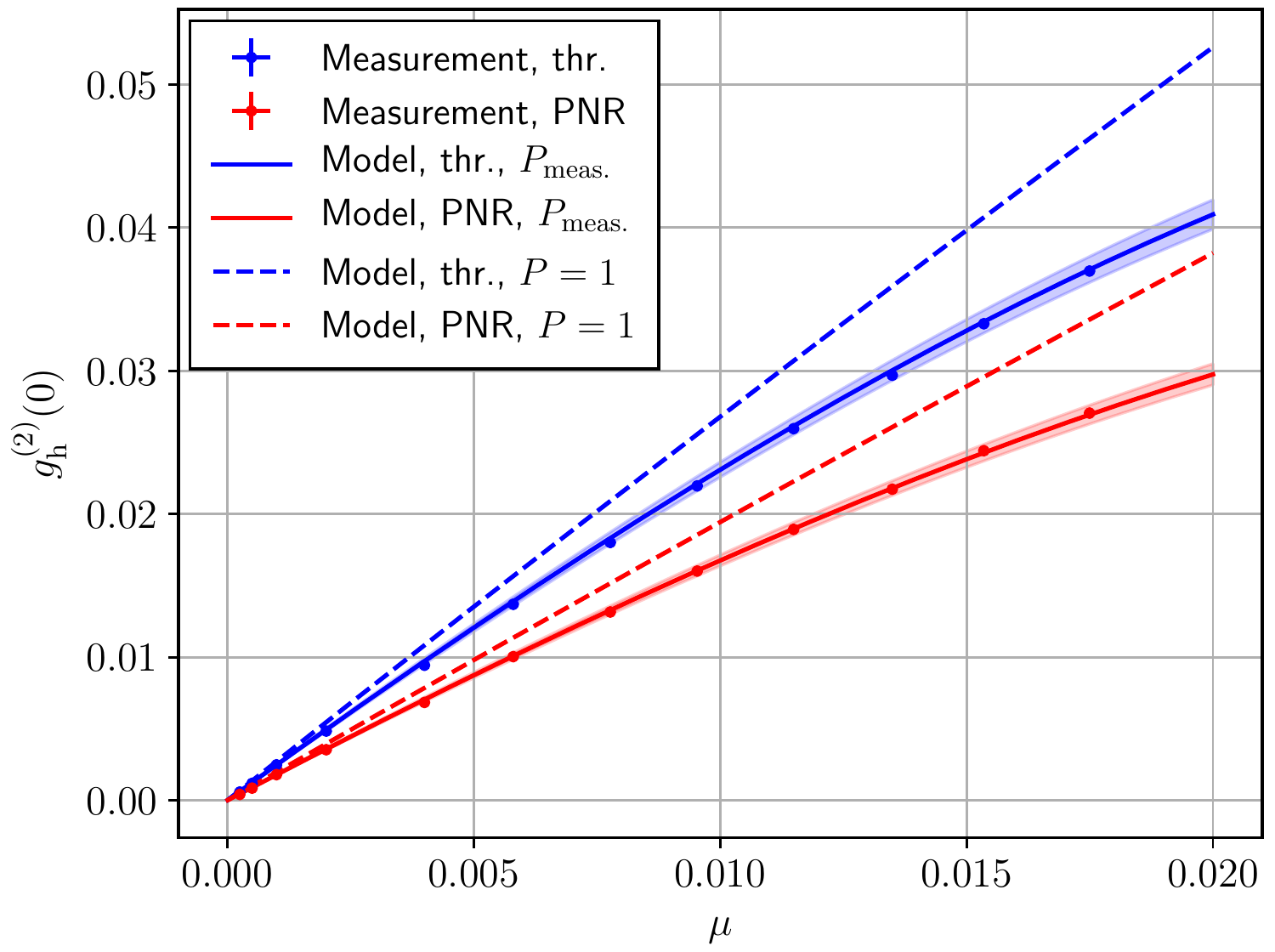}
    \caption{Heralded second-order autocorrelation function as a function of the mean photon number, $\mu$. 
    The blue points correspond to the case where the {P-SNSPD} operates in threshold configuration (thr.), %($n\geq 1$) 
    whereas the red points show the measurements for PNR configuration. 
    %the single-photon heralding mode. %($n=1$).
    The solid lines are obtained from the theoretical model with the same purity as in the experiment, where the shaded areas mark the spectral purity interval of $\pm\SI{4}{\%}$. The dashed lines show the behavior for a source with purity $P=1$.}
    \label{fig:g2}
\end{figure}

The results of the first experiment (see \cref{fig:setup}a), where we use the \psnspd{} as the heralding detector for the HSPS, are shown in \cref{fig:g2}. We calculate the mean photon number $\mu$ from the measured probability of detecting a heralding photon, $p_\mathrm{h}$, in the threshold configuration together with the characterized total efficiency of the heralding photon $\eta_\mathrm{h}=(C_\mathrm{ha}+C_\mathrm{hb})/(C_\mathrm{a}+C_\mathrm{b})=\num{0.6348(5)}$ and the Schmidt coefficients $\lambda_k$ obtained from a fit of \cref{eq:g2uncond} to the corresponding measured data (see \cref{eq:ph} in Appendix~\ref{sec:app_model}). %($n\geq 1$). 
The values for $g_{\mathrm{h}}^{(2)}(0)$ are calculated according to \cref{eq:g2cond}, where the blue points correspond to the threshold configuration %($n\geq 1$) 
of the \psnspd{} and the red points to the PNR configuration. %($n=1$). 
The solid lines are obtained from the theoretical model described in \cref{sec:theory} with the characterized efficiencies  $\eta_a=2 C_\mathrm{ha}/C_\mathrm{h}=\num{0.6293(6)}$ and $\eta_b=2  C_\mathrm{hb}/C_\mathrm{h}=\num{0.5809(6)}$ as well as $\eta_\mathrm{h}$ and $\lambda_k$ as above. The shaded areas around the solid lines mark the regions with a difference in spectral purity of $\pm\SI{4}{\%}$ and the dashed lines show the theory calculation for a spectrally pure source. 

The ratio between the blue and the red data is $g_{\mathrm{h,thr}}^{(2)}(0)/g_{\mathrm{h,PNR}}^{(2)}(0)$ which is the factor by which the heralding rate can be increased when switching from threshold to PNR heralding mode while keeping a fixed $g_{\mathrm{h}}^{(2)}(0) \ll 1$. This inversely corresponds to a reduction in $g_{\mathrm{h}}^{(2)}(0)$ of \SI{26.6(2)}{\%}. It should be noted that, in this demonstration, no spectral filtering of the heralding photons has been performed, in order to show the full potential of the PNR detector in terms of reduction of $g_{\mathrm{h}}^{(2)}(0)$.

\begin{figure}
    \capstart
    \centering
    \includegraphics[width= 1\columnwidth]{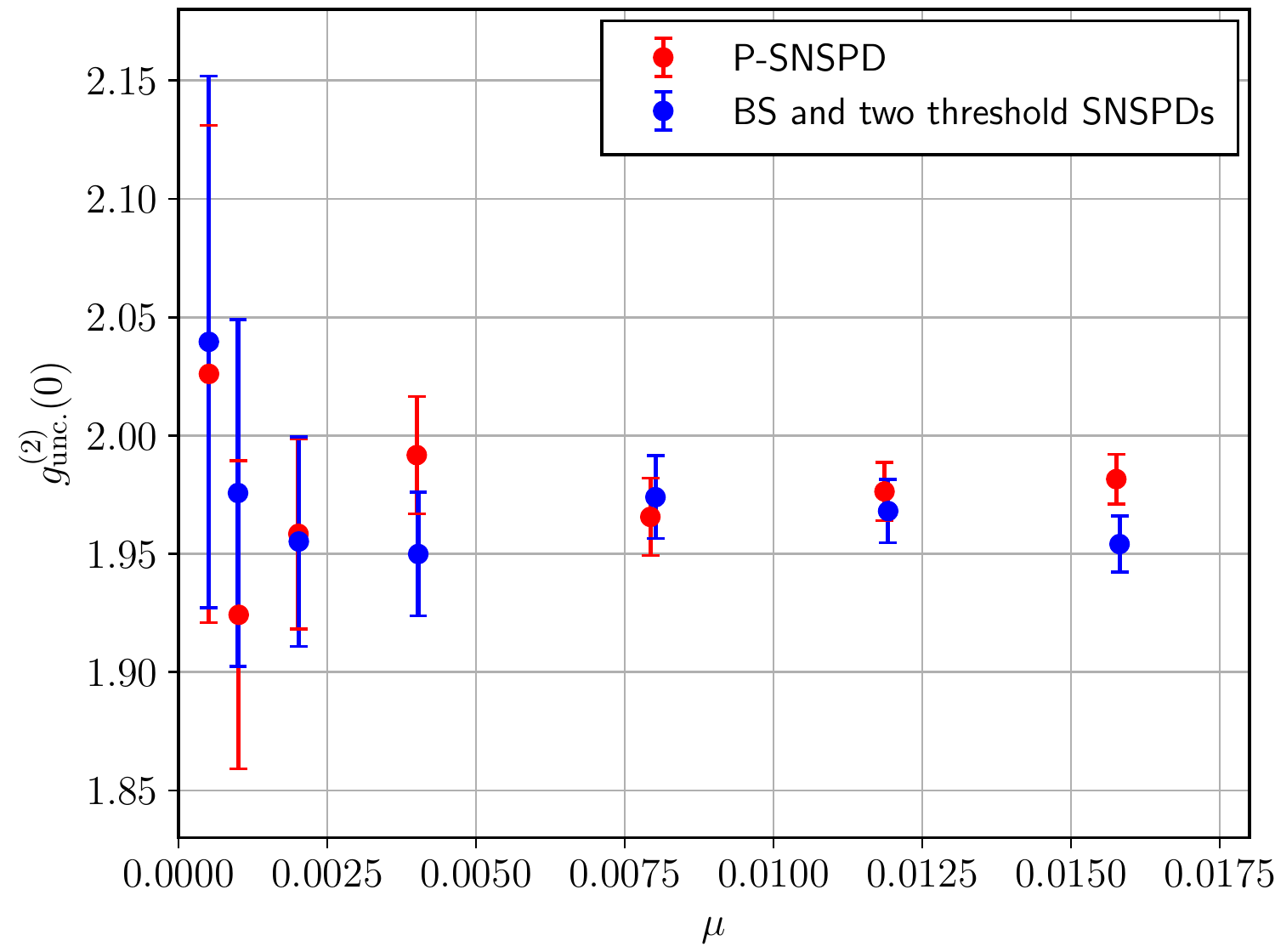}
    \caption{Unconditional second-order autocorrelation measurement on the spectrally filtered idler mode of the SPDC source. The red data correspond to the measurement with the {P-SNSPD}, whereas the blue data are obtained with the standard method using a beam splitter and two threshold detectors.}
    \label{fig:g2unc}
\end{figure}

As described in \cref{sec:experiment}, the second experiment (see \cref{fig:setup}b) we reconstruct the photon-number probability distribution of one mode of the TMSV state by measuring the click probability distribution with the \psnspd{}. We then calculate $g_\mathrm{unc.}^{(2)}(0)$ using \cref{eq:g2} and compare it to the value obtained by the standard method using \cref{eq:g2uncond}. The results for different mean photon numbers are shown in Fig.~\ref{fig:g2unc}. 

The error bars for the \psnspd{} were calculated through a Monte Carlo method with $10^3$ iterations. 
In each iteration, to characterize \textbf{P}, the Poissonian input state used for the detector characterization is randomly picked from a Gaussian distribution centered at $\mu =1$ (the experimental value) with a standard deviation of $\sigma=0.05$. In this way, we take into account the uncertainties of our characterization setup (see Ref. \cite{stasi2022high} for more details).
The obtained matrix \textbf{P} is then used to reconstruct the light input statistics $p_m$ of the SPDC source from the experimental photon-counting distribution of the \psnspd{}. As a last step in each iteration, the value  $g_{\mathrm{unc.}}^{(2)}(0)$ is computed from the reconstructed statistics using \cref{eq:g2}. 

%%%%%%%%%%%%%%%%%%%%%%%%%%%%%%%%%%%%%%%%%%%%%%%%%%%%%%%%%%%%%%%%%%%%%%%%%%%%%%%%%%%%%%%%%%%%%%%%%%%%%%%%%%%%%%%%%%%%%%%%%%%%%%%%%%%%%%%%%%%%%%

\section{Discussion}
\label{sec:discussion}
In order to improve the HSPS, the most important parameters to optimize are the PNR capability of the detector and the overall efficiency of the heralding path $\eta_\mathrm{h}$, i.e. transmittance through optical elements, coupling and detection efficiency~\cite{Sempere-Llagostera2022}. For a pure SPDC source combined with a perfect PNR heralding detector ($\mathbf{P}=\id$), the ratio $g_{\mathrm{h,thr}}^{(2)}(0)/g_{\mathrm{h,PNR}}^{(2)}(0)$ diverges for $\eta_\mathrm{h}\rightarrow 1$. It reaches a value of 3 for $\eta_\mathrm{h}=0.8$ and surpasses 10 for $\eta_\mathrm{h}=0.95$. For the total efficiency measured in our experiment $\eta_\mathrm{h}=0.635$, a perfect PNR detector would achieve $g_{\mathrm{h,thr}}^{(2)}(0)/g_{\mathrm{h,PNR}}^{(2)}(0)=1.87$, however, our reported value of \num{1.363(6)} lies significantly lower. This is due to the fact that the PNR capability of the detector to correctly detect an incoming higher-photon-number state is still limited by the non-resolvability of two photons hitting the same pixel. In order to increase the PNR capability of our \psnspd{}, the number of pixels needs to be increased while still maintaining good amplitude discrimination of the electrical readout signal between different photon-number detection events. In addition, a uniform light distribution over the pixels can be obtained by exploiting an interleaved design, which reduces the probability of having more photons arriving on the same pixel \cite{zhang201916}. %Work is currently ongoing into improving the \psnspd{} through such design changes.

Another important aspect of a HSPS is the spectral purity of the heralded photon, since many single-photon applications relying on interference require high purity. The standard approach to increase the purity is to spectrally filter the heralding photons. This presents a viable solution also in the case of a PNR heralding detector, as long as the insertion loss of the spectral filter is sufficiently low, as discussed above. In our demonstration, %of the PNR HSPS 
we did not spectrally filter the heralding photons in order to show the full potential of the \psnspd{} in terms of reduction of $g_{\mathrm{h}}^{(2)}(0)$. 
For increasing pump power, we observe a decrease in purity due to spectral broadening of the pump light caused by the non-linearity of the spatial mode cleaning fiber (Coherent 780-HP, \SI{9}{cm} long) before the PPKTP crystal. We take this into account for our theory model by fitting \cref{eq:g2uncond} to the experimental data with the purity $P$ as a fit parameter, see Appendix~\ref{sec:app_model}. If the source needs to be operated at a pump power $\geq \SI{50}{mW}$ ($\mu\geq 0.003$), this issue could be solved by replacing the standard fiber for spatial mode cleaning by a photonic crystal fiber, where nonlinear effects are largely suppressed.

Lastly, we showed that \psnspd{}s can perform unconditional  $g^{(2)}(0)$ measurements, replacing the standard method where two threshold detectors are needed. As it can be seen from the results, the $g_{\mathrm{unc.}}^{(2)}(0)$ obtained with the two different methods match very well, both in the absolute value and in the amplitude of the error bars. This result demonstrates that \psnspd{}s can effectively replace a 50/50 beam splitter and two detectors as used in the standard method, simplifying the overall experimental apparatus.
It is important to note that the approach does not rely on an accurate absolute characterization of the \psnspd{} efficiency, keeping the $g_\mathrm{unc.}^{(2)}(0)$ measurement efficiency-independent as in the case of the more common-place method which uses two detectors. However, this is not the case for the reconstructed photon-number distribution, $p_n$.

%%%%%%%%%%%%%%%%%%%%%%%%%%%%%%%%%%%%%%%%%%%%%%%%%%%%%%%%%%%%%%%%%%%%%%%%%%%%%%%%%%%%%%%%%%%%%%%%%%%%%%%%%%%%%%%%%%%%%%%%%%%%%%%%%%%%%%%%%%%%%

\section{Conclusion}
\label{sec:conclusion}
We have shown the benefit for a HSPS that a PNR detector, such as a \psnspd{}, can bring when combined with a SPDC source. Even though \psnspd{}s do not possess perfect PNR capability, we already demonstrate a significant reduction in the heralded $g^{(2)}(0)$ of \SI{26.6(2)}{\%} when the detector is operated in PNR configuration, compared to the threshold configuration. In addition, the PNR capability of the \psnspd{} is not limited by the timing jitter as in the case of a slew rate discrimination, or by super-short light pulses (tens~of~ps) as in the taper approach. Lastly, to improve our results further, the number of pixels of the \psnspd{} need to be increased. In that case, not only a better photon-number discrimination can be achieved, but also an overall faster recovery time, thanks to the \psnspd{} structure itself.
This result marks a first step towards high-rate generation of single photons which could be of use in repeater protocols~\cite{Sangouard2007}. 
We further show the usefulness of our \psnspd{} for the task of reconstructing the photon-number probability distribution of pulsed light by measuring a thermal state. We demonstrated that we can estimate $g^{(2)}(0)$ with a single detector, which replaces a beam splitter and two threshold detectors used in more common-place methods.
Therefore, such PNR detectors can be of great help in quantum metrology applications and source characterization.
% To improve our results further, the number of pixels of the \psnspd{} as well as the single-pixel detection efficiencies need to be increased, which would yield better photon-number discrimination. 

%%%%%%%%%%%%%%%%%%%%%%%%%%%%%%%%%%%%%%%%%%%%%%%%%%%%%%%%%%%%%%%%%%%%%%%%%%%%%%%%%%%%%%%%%%%%%%%%%%%%%%%%%%%%%%%%%%%%%%%%%%%%%%%%%%%%%%%%%%%%%%

\begin{acknowledgments}
We thank Giovanni V. Resta and Gaëtan Gras for useful discussion.
This work was supported by the Swiss National Science Foundation SNSF (Grant No.~200020\_182664). L.S. is part of the AppQInfo MSCA ITN which received funding from the EU Horizon 2020 research and innovation program under the Marie Sklodowska-Curie grant agreement No. 956071.
\end{acknowledgments}
\medskip
L.S. and P.C. contributed equally to this work.

%%%%%%%%%%%%%%%%%%%%%%%%%%%%%%%%%%%%%%%%%%%%%%%%%%%%%%%%%%%%%%%%%%%%%%%%%%%%%%%%%%%%%%%%%%%%%%%%%%%%%%%%%%%%%%%%%%%%%%%%%%%%%%%%%%%%%%%%%%%%%%
\medskip
\appendix
\section{Theoretical model}
\label{sec:app_model}
We use a characteristic function based approach to model our SPDC source and to derive the single and coincidence detection probabilities~\cite{Takeoka2015}. We start with the covariance matrix of a TMSV state and apply Gaussian unitary operations corresponding to the action of the beam splitters in our model as shown in Fig.~\ref{fig:model}. On each mode, we apply a loss channel with the corresponding transmittance $\eta_\mathrm{a},\eta_\mathrm{b}$ and $\eta_\mathrm{h}$. Each detector is described by a positive operator valued measure (POVM) with element $E_0=\ketbra{0}{0}^{\otimes M}$ corresponding to a no-click outcome and $E_c=\id - \ketbra{0}{0}^{\otimes M}$ corresponding to a click outcome over all the spectral modes $M$. This leads to the following detection probability of a detecting a heralding photon per pump pulse
%figure was here with 0.8\columnwidht

\begin{widetext}
% \onecolumngrid

\begin{figure}[h]
\capstart
\includegraphics[width = 0.5\columnwidth]{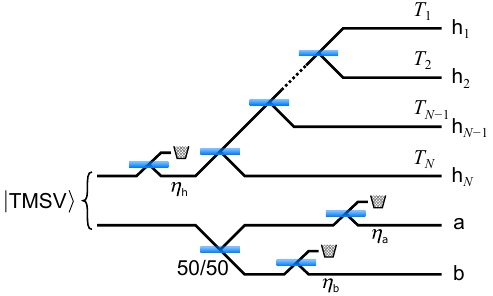}
\caption{\label{fig:model} 
Schematic representation of the theory model to calculate the single and coincidence detection probabilities. The heralding mode of the two-mode squeezed vacuum (TMSV) state is subject to loss ($\eta_\mathrm{h}$) and is then split into $N$ modes before being detected by threshold detectors ($\mathrm{h}_1$ to $\mathrm{h}_N$) on each mode.
The other mode of the TMSV state is sent to a 50/50 beam splitter and further undergoes loss channels ($\eta_\mathrm{a}$ and $\eta_\mathrm{b}$) before reaching the threshold detectors $\mathrm{a}$ and $\mathrm{b}$.}
\end{figure}

\begin{equation}
\label{eq:ph}
    p_\mathrm{h} = \sum_{k=1}^N \tr{\rho \, \big(\id_\mathrm{a} \otimes \id_\mathrm{b} \otimes E_{c,\mathrm{h}_k} \otimes E_{0,\mathrm{h}_{\neg k}}^{\otimes (N-1)}\big)} 
    = \sum_{k=1}^N \bigg(\prod_{m} \frac{1}{1+(1-T_k)\eta_\mathrm{h} \lambda_m \mu}\bigg) - N \prod_{m} \frac{1}{1+\eta_\mathrm{h} \lambda_m \mu}, 
\end{equation}
where $\{\lambda_m\}_{m=1}^{M}$ denote the Schmidt coefficients (with $\sum_{m}\lambda_m=1$) and $T_k$ is the fraction of the light that reaches pixel $k$ of the heralding detector with $\sum_k T_k=1$. Note that here and in the following, taking $\{T_k\}_{k=1}^{N}$ models a single-photon detection event on the \psnspd{}, but by setting $N=1$ and $T_1=1$ one obtains the behavior for the detector in threshold configuration, i.e. detecting one or more photons. 

Similarly to $p_\mathrm{h}$, we obtain the probability of a coincidence detection between the heralding detector and detector $\mathrm{a}$ after the 50/50 beam splitter on the heralded mode
\begin{equation}
\label{eq:pha}
\begin{split}
    p_{\mathrm{ha}} &= \sum_{k=1}^N \tr{\rho \, \big(E_{c,\mathrm{a}} \otimes \id_\mathrm{b} \otimes E_{c,\mathrm{h}_k} \otimes E_{0,\mathrm{h}_{\neg k}}^{\otimes (N-1)}\big)} \\
    &= p_\mathrm{h} - \sum_{k=1}^N \bigg(\prod_{m} \frac{2}{2+[(1-T_k)(2-\eta_\mathrm{a})\eta_\mathrm{h} +\eta_\mathrm{a}] \lambda_m \mu}\bigg) 
    + N \prod_{m} \frac{2}{2+[(2-\eta_\mathrm{a}) \eta_\mathrm{h} +\eta_\mathrm{a}] \lambda_m \mu}
\end{split}
\end{equation}
and the three-fold coincidence probability
\begin{equation}
\label{eq:phab}
\begin{split}
    p_{\mathrm{hab}} &= \sum_{k=1}^N \tr{\rho \, \big(E_{c,\mathrm{a}} \otimes E_{c,B} \otimes E_{c,\mathrm{h}_k} \otimes E_{0,\mathrm{h}_{\neg k}}^{\otimes (N-1)}\big)} \\
    &= p_\mathrm{ha}+p_\mathrm{hb}-p_\mathrm{h} + \sum_{k=1}^N \bigg(\prod_{m} \frac{2}{2+[(1-T_k)(2-\eta_\mathrm{a}-\eta_\mathrm{b})\eta_\mathrm{h} +\eta_\mathrm{a}+\eta_\mathrm{b}] \lambda_m \mu}\bigg) \\
    &\quad - N \prod_{m} \frac{2}{2+[(2-\eta_\mathrm{a}-\eta_\mathrm{b}) \eta_\mathrm{h} +\eta_\mathrm{a}+\eta_\mathrm{b}] \lambda_m \mu}.
\end{split}
\end{equation}
The detection probability of detector $\mathrm{a}$ (and similarly for detector $\mathrm{b}$) is given by 
\begin{equation}
\label{eq:pa}
    p_\mathrm{a} = \sum_{k=1}^N \tr{\rho \, \big(E_{c,\mathrm{a}} \otimes \id_\mathrm{b} \otimes \id_\mathrm{h}^{\otimes N} \big)} 
    = 1 - \prod_{m} \frac{1}{1+\eta_\mathrm{a} \lambda_m \mu}
\end{equation}
and the coincidence probability between detector $\mathrm{a}$ and $\mathrm{b}$ by
\begin{equation}
\label{eq:pab}
    p_\mathrm{ab} = \sum_{k=1}^N \tr{\rho \, \big(E_{c,\mathrm{a}} \otimes E_{c,\mathrm{b}} \otimes \id_\mathrm{h}^{\otimes N} \big)} 
    = 1 - \prod_{m} \frac{2}{2+\eta_\mathrm{a} \lambda_m \mu} - \prod_{m} \frac{2}{2+\eta_\mathrm{b} \lambda_m \mu} + \prod_{m} \frac{2}{2+(\eta_\mathrm{a}+\eta_\mathrm{b}) \lambda_m \mu}.
\end{equation}
Those formulas are then used to calculate the theory values for the second-order autocorrelation functions according to Eqs.~\eqref{eq:g2cond} and \eqref{eq:g2uncond}. In Fig.~\ref{fig:prob_measurement}, the theory model for all the different probabilities is shown together with the experimental data. 

\begin{figure}
\capstart
\includegraphics[width = \columnwidth]{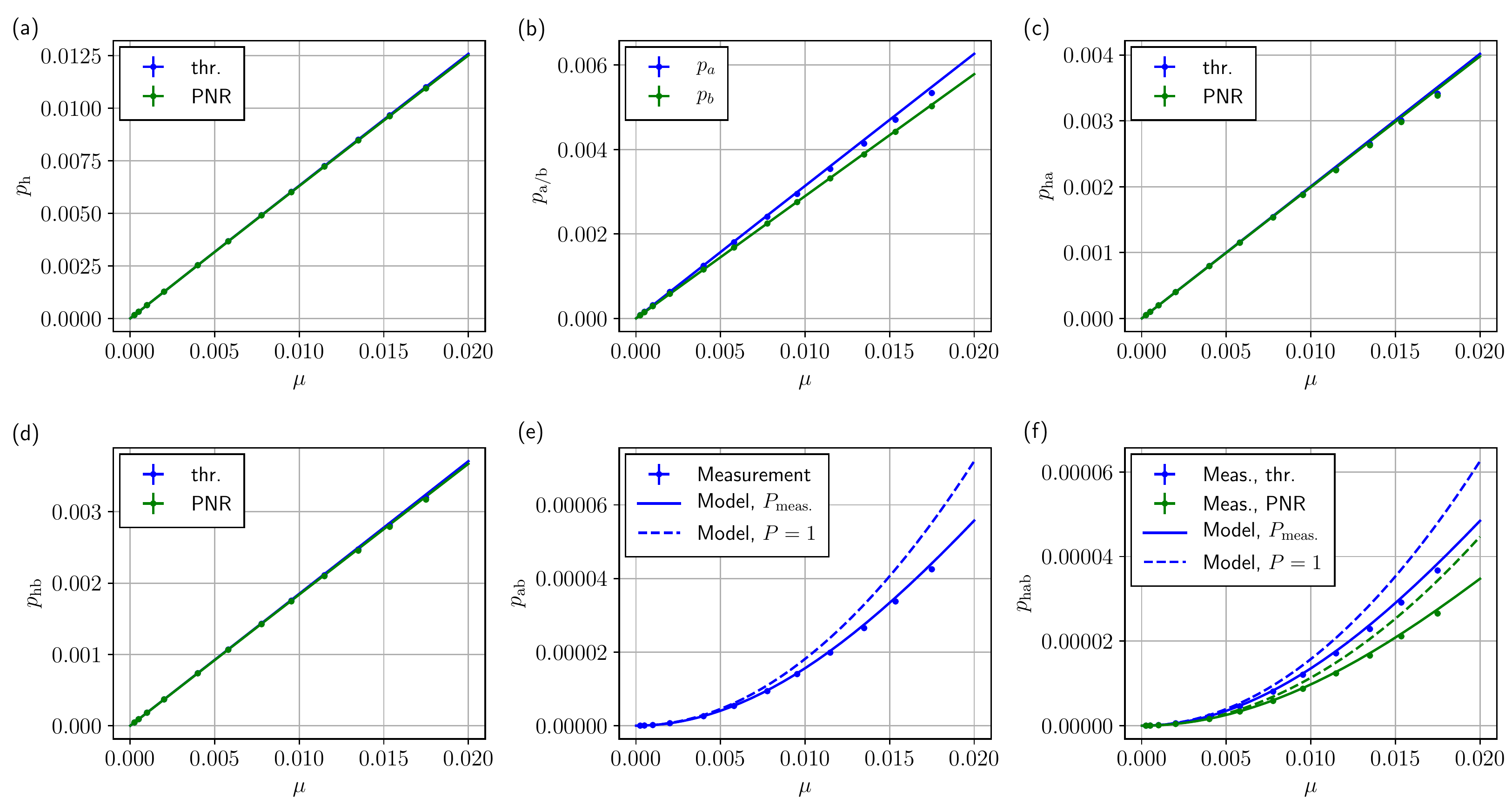}
\caption{\label{fig:prob_measurement} 
Measurement and theory model for the single and coincidence detection probabilities. The solid lines are obtained from the theory model including the non-unit purity of our SPDC source whereas the dashed lines describe the behavior for a pure source. (a) Probability of a heralding detection per pump pulse for the threshold and PNR configurations. (b) Detection probabilities for detectors $\mathrm{D_a}$ and $\mathrm{D_b}$. (c) Coincidence probability between detectors $\mathrm{D_h}$ and $\mathrm{D_a}$. (d) Coincidence probability between detectors $\mathrm{D_h}$ and $\mathrm{D_b}$. (e) Coincidence probability for detections on $\mathrm{D_a}$ and $\mathrm{D_b}$. (f) Triple-coincidence probability between detectors $\mathrm{D_h}$, $\mathrm{D_a}$ and $\mathrm{D_b}$.}
\end{figure}

\end{widetext}
% \twocolumngrid

For a spectrally pure source (i.e. $\lambda_1=1$), the derived formulas (\ref{eq:ph}-\ref{eq:pab}) simplify. In the case of non-unit spectral purity $P=\sum \lambda_k^2 <1$, but still $P>\frac{1}{2}$, we estimate $P$ by fixing two Schmidt coefficients $\lambda_{1,2} = \frac{1}{2}\pm \sqrt{\frac{P}{2}-\frac{1}{4}}$ and fitting $g_\mathrm{unc.}^{(2)}(0)\approx p_\mathrm{ab}/p_\mathrm{a}p_\mathrm{b}$ to the experimental value. In our case, we find that the purity of the source decreases as a function of $\mu$, therefore a second-order polynomial is fitted to the purity values obtained from $g_\mathrm{unc.}^{(2)}(0)$ to get the function $P_\mathrm{meas.}(\mu)$ which is then used throughout in the theoretical model. 

We attribute the decrease in purity for increasing pump power to the nonlinearity of the spatial mode cleaning fiber in our setup before the nonlinear crystal. This behavior is confirmed by the measurement of the pump spectrum as a function of pump power and a simulation of the corresponding joint spectral amplitude of the down-converted photon pairs~\cite{Guerreiro2013,Ljunggren2005,Zielnicki2018}.

%%%%%%%%%%%%%%%%%%%%%%%%%%%%%%%%%%%%%%%%%%%%%%%%%%%%%%%%%%%%%%%%%%%%%%%%%%%%%%%%%%%%%%%%%%%%%%%%%%%%%%%%%%%%%%%%%%%%%%%%%%%%%%%%%%%%%%%%%%%%%%

%\bibliographystyle{apsrev4-2}
%\bibliographystyle{unsrtnat}
\bibliographystyle{ieeetr}
\bibliography{./references}

\end{document}